\documentclass[10pt,a4paper,twocolumn,pra,superscriptaddress,amsmath,amssymb,
               mathrsfs,showpacs]{revtex4-1}
\usepackage{tabularx}
\usepackage{epstopdf,graphicx}
\usepackage{braket}
\usepackage{units}
\usepackage{upgreek}
\usepackage{xcolor}
\usepackage{hyperref}
\hypersetup{%
    colorlinks,
    linkcolor={black},
    citecolor={black},
    urlcolor={black}}
\usepackage{mathtools}
\DeclarePairedDelimiter\abs{\lvert}{\rvert}

\newcommand{\Vpcm}[1]{\unit[#1]{V/cm}}
\newcommand{\mVpcm}[1]{\unit[#1]{mV/cm}}
\newcommand{\MHz}[1]{\unit[#1]{MHz}}
\newcommand{\ns}[1]{\unit[#1]{ns}}

\begin{document}

\title{Fast and accurate circularization of a Rydberg atom}

\author{Sabrina Patsch}
\affiliation{Theoretische Physik, Universit\"{a}t Kassel,
Heinrich-Plett-Stra{\ss}e 40, D-34132 Kassel, Germany}

\author{Daniel M. Reich}
\affiliation{Theoretische Physik, Universit\"{a}t Kassel,
Heinrich-Plett-Stra{\ss}e 40, D-34132 Kassel, Germany}

\author{Jean-Michel Raimond}
\affiliation{Laboratoire Kastler Brossel, Coll\`ege de France,
  CNRS, ENS-Universit\'e PSL,
  Sorbonne Universit\'e}

\author{Michel Brune}
\affiliation{Laboratoire Kastler Brossel, Coll\`ege de France,
  CNRS, ENS-Universit\'e PSL,
  Sorbonne Universit\'e}

\author{S\'{e}bastien Gleyzes}
\affiliation{Laboratoire Kastler Brossel, Coll\`ege de France,
  CNRS, ENS-Universit\'e PSL,
  Sorbonne Universit\'e}

\author{Christiane P. Koch}
\affiliation{Theoretische Physik, Universit\"{a}t Kassel,
Heinrich-Plett-Stra{\ss}e 40, D-34132 Kassel, Germany}
\email{christiane.koch@uni-kassel.de}
\hyphenation{Ryd-berg sen-sing ma-ni-fold}

\date{\today}
\begin{abstract}
  Preparation of a so-called circular state in a Rydberg atom where the
  projection of the electron angular momentum takes its maximum value is
  challenging due to the required amount of angular momentum transfer. Currently
  available protocols for circular state preparation are either accurate but
  slow or fast but error-prone. Here, we show how to use quantum optimal control
  theory to derive pulse shapes that realize fast and accurate circularization
  of a Rydberg atom. In particular, we present a theoretical proposal for
  optimized radio-frequency pulses that achieve high fidelity in the shortest
  possible time, given current experimental limitations on peak amplitudes and
  spectral bandwidth. We also discuss the fundamental quantum speed limit for
  circularization of a Rydberg atom, when lifting these constraints.
\end{abstract}

\maketitle

\section{Introduction}
\label{sec:intro}

Circular Rydberg levels are quantum states of the valence electron that are
characterized by a large principal quantum number and a maximum angular momentum
projection \cite{Hulet1983}. They can be prepared by optical excitation of
ground-state atoms into a low-angular-momentum Rydberg state, which is then
exposed to a DC field, lifting the degeneracy of the Stark manifold, and to
a near-resonant radio-frequency field, providing the required angular
momentum~\cite{Signoles2017}.

Their long lifetime makes these states an ideal tool for applications in quantum
technology. For example, they provide a key ingredient for microwave cavity
quantum electrodynamics (CQED)~\cite{RaimondRMP01}, enabling the  generation of
non-classical states of a cavity mode~\cite{BrunePRL96}. Recently, they have
attracted attention in the context of quantum interfaces~\cite{MaxwellPRA14} and
quantum-enhanced sensing and metrology~\cite{Facon2016,RamosPRA17}.  In more
detail, an electrometer with record sensitivity has been demonstrated using
a superposition of two circular states in adjacent Stark manifolds of a rubidium
Rydberg atom~\cite{Facon2016}. The repetition rate of the experiment is
ultimately limited by the time required for the circular state preparation.
Sensing of magnetic, instead of electric, fields, would be enabled by creating
a coherent superposition of two circular states with opposite angular momentum
projection quantum numbers. Both the electrometer and the magnetometer require
the state preparation to proceed sufficiently fast to beat unavoidable
decoherence.

Similarly, using Rydberg atoms to build an interface between optical and
microwave photons~\cite{MaxwellPRA14} relies on fast coherent transfer between
low- and high-angular-momentum states: While low-angular-momentum Rydberg states
couple to optical photons and thus have short lifetimes, circular states do not.
Circular states do, however, couple strongly to microwave photons.  In other
words, it is circularization from low- to high-angular-momentum states, together
with its inverse process, that provides the link to interface optical and
microwave photons. However, such an interface will work reliably only when the
transfer proceeds both sufficiently fast and with high accuracy.

Fast coherent transfer to the circular state has recently been demonstrated by
coupling a Rydberg atom to a radio-frequency (RF) field with a well-defined
polarisation~\cite{Signoles2017}. The near-resonant RF drive implements a Rabi
oscillation between a low-angular-momentum state and the circular one, which
allows the circularization to proceed in about 200$\,$ns. However, the transfer
rate was limited to about 80\% by the anharmonicity of the Stark manifold and
perturbations due to the finite quantum defects of the low-angular-momentum
states~\cite{Signoles2017}.  Another possibility to perform the circularization
is rapid adiabatic passage~\cite{Hulet1983, Nussenzveig1993}, which relies on
the slow transformation of instantaneous eigenstates using chirped RF pulses or
slow variations of the DC field.  Here, fidelities close to  100\% are
achievable, but the required time is much longer, namely several microseconds.
As a result, large dynamic phases are accumulated which are error-prone due to
imperfections in the control. These errors propagate in the microwave-to-optical
interface or when generating a coherent superposition of opposite angular
momentum states. This problem of adiabatic passage is generic; it is also
encountered, for example, when utilizing stimulated Raman adiabatic passage
(STIRAP) to realize quantum gates in Rydberg atoms: While population can be
transferred very efficiently, precise control over the phase is extremely
difficult and the famous robustness of STIRAP is lost~\cite{GoerzPRA14}. In the
context of quantum-enhanced sensing and metrology or quantum interfaces,
adiabatic passage thus does not provide a viable route.

Further improvement of the coherent transfer protocol of
Ref.~\cite{Signoles2017} is hampered by the complexity of the dynamics that
proceeds in a comparatively large Hilbert space.  Fast and accurate preparation
of a circular Rydberg state thus remains an open challenge. Here, we show how to
use quantum optimal control theory to tackle this problem.

Quantum optimal control theory is based on defining a figure of merit, here the
state preparation fidelity, and treating it as a functional of the external
controls.  The latter are then determined using e.g. variational
calculus~\cite{Glaser2015}.  The control problem is most often solved
numerically, and large Hilbert spaces do not pose a problem as long as the time
evolution of the system can be calculated with reasonable numerical resources.
Quantum optimal control theory is an ideal tool whenever high fidelity is
desired~\cite{Glaser2015}. Being biased against adiabatic
solutions~\cite{YuanPRA12}, it is, moreover, well suited to identify protocols
that require the minimum amount of time for a given process~\cite{Caneva2009}.
For example, quantum optimal control theory has been used to determine the
shortest time required for transport in a spin chain~\cite{Caneva2009} and the
fastest universal set of gates within circuit QED~\cite{GoerzNPJQI17}.

The paper is organized as follows. Section~\ref{sec:model} introduces our model
to describe the Rydberg atom, whereas quantum optimal control theory is briefly
reviewed in Sec.~\ref{sec:oct}. Section~\ref{sec:exp_circ} explains how, based
on a thorough understanding of the circularization dynamics exploited in
Ref.~\cite{Signoles2017}, a guess pulse is constructed that, when optimized,
provides the desired fast and accurate transfer while obeying current
experimental constraints. The robustness of our control solution is analyzed for
various noise sources in Sec.~\ref{sec:noise}.  Finally, Sec.~\ref{sec:qsl}
explores how far the quantum speed limit can be pushed when lifting typical
experimental constraints, and Sec.~\ref{sec:concl} concludes.

\section{Model}
\label{sec:model}

  We consider the single valence electron of an alkali atom.  Without an
  external field, the energy spectrum is given by the quantum defect
  theory~\cite{Seaton1983}.  Briefly, a complete set of quantum numbers consists
  of the principal, orbital angular momentum, and projected angular momentum
  quantum numbers, $n$, $\ell$ and $m_\ell$, similarly to the hydrogen atom.
  Imperfect shielding of the nuclear charge by the core electrons breaks the
  degeneracy of the eigenenergies for small $\ell$, when the valence electron is
  close to the nucleus. The corresponding quantum defect is accounted for by
  a correction $\delta_{n \ell j}$ to the Rydberg formula for the
  eigenenergies~\cite{Gallagher}
  \begin{equation}
    E_{n \ell j} = - \frac{1}{2(n-\delta_{n \ell j})^2},
    \label{eq:E_nl}
  \end{equation}
  where atomic units ($\hbar=1$) are employed throughout and $j$ is the total
  angular momentum quantum number, $\abs{\ell-s} \leq j \leq \ell + s$.  Since
  the quantum defect does not break the spherical symmetry, $n$, $\ell$, and
  $m_\ell$ are still good quantum numbers.  In our calculations, we use
  a perturbative expansion of $\delta_{n \ell j}$ up to second order. We use the
  values given in Ref.~\cite{Signoles2014}, corresponding to rubidium $85$ for
  $n \geq 20$. The quantum defect is neglected for states with $\ell>7$ and we
  neglect spin-orbit coupling. Because the quantum defect is $j$-dependent, we
  choose $j=\ell+s=\ell+\tfrac{1}{2}$ throughout.

  \begin{figure}[tb]
    \centering
    \includegraphics{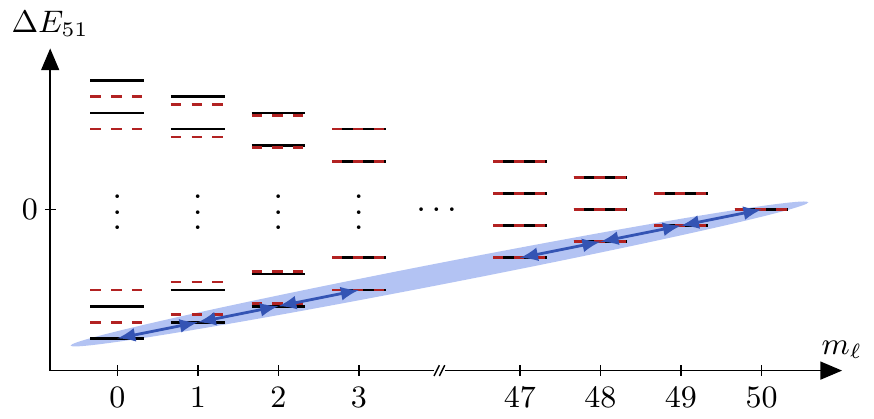}
    \caption{Schematic spectrum for $m_\ell \geq 0$ of the $n=51$-manifold for
      hydrogen (black) and rubidium (dashed red) with a moderate DC field. The
      zero of energy was set to the field-free position of the manifold. The
      states of rubidium with $\abs{m_\ell}\leq 2$ are significantly affected by
      the quantum defect. The blue ellipse indicates the lowest diagonal ladder
      whose states are connected by $\sigma^+$-polarized transitions (blue
      arrows).
    }
    \label{fig:levels}
  \end{figure}

  In the presence of a DC electric field, the spherical symmetry is broken and
  $\ell$ is not a good quantum number anymore. However, the component of the
  angular momentum along the direction of the DC field is conserved such that
  $m_\ell$ is still a good quantum number. Incidentally, neglecting the quantum
  defect (i.e.\ considering a hydrogen atom), the projection of the Runge-Lenz
  vector along the direction of the DC field is also conserved. This gives rise
  to the eccentricity quantum number $\mu$, which, for given $n$ and $m_\ell$,
  takes the values
    $-\left(n-\abs{m_{\ell}}-1\right),
     -\left(n-\abs{m_{\ell}}-1\right)+2,
      \dots,
      \left(n-\abs{m_{\ell}}-1\right)-2,
      \left(n-\abs{m_{\ell}}-1\right)$.
  Note that $\mu$ alternates between only odd and only even numbers for
  different values of $m_\ell$.  Perturbation theory in first and second order
  yields the DC Stark shifts~\cite{Mur1993},
  \begin{subequations}
    \label{eq:DeltaE_DC}
    \begin{eqnarray}
      \Delta E^{\left(1\right)}
        & = & \frac{3}{2}\mu\, n\,\mathcal E_\text{DC}\,,\label{eq:DeltaE_DC1}\\
      \Delta E^{\left(2\right)}
        & = & -\frac{1}{16}n^{4} \left(17n^{2}-3\mu^{2}-9m_{\ell}^{2}+19\right)
              \mathcal E_\text{DC}^{2}\,,\label{eq:DeltaE_DC2}
    \end{eqnarray}
  \end{subequations}
  where $\mathcal E_\text{DC}$ denotes the DC field strength, which is taken to
  be of the order of $\Vpcm{1}$. For such moderate DC field strengths,  the
  second-order Stark effect can be neglected and the eigenvalues form ladders
  for fixed $n$ and $m_\ell$ (cf.\ black energy levels in
  Fig.~\ref{fig:levels}). The ladders with $\pm m_\ell$ are identical, and
  neighboring ladders, $\Delta m_\ell=\pm1$, have an energy offset of half
  a ladder step. The number of steps in each ladder is given by
  $n-\abs{m_\ell}$.

  An RF field,  perpendicular to the DC one, induces transitions between
  neighboring ladders.  Pulses with a $\sigma^+$ polarization drive transitions
  between states with $\Delta m_\ell=+1$ whilst increasing the energy, whereas
  transitions between states with $\Delta m_\ell=-1$ are accompanied by
  a decrease in energy.  The opposite rules apply for $\sigma^-$-polarized
  pulses.  The states where the magnetic and the eccentricity quantum number
  change by one lie on so-called diagonal ladders.  In the following, we will
  only consider the lowest diagonal on the right hand side of the manifold, with
  $m_\ell \geq 0$. Its states are connected by $\sigma^+$-polarized transitions
  (cf.\ blue ellipse and arrows in Fig.~\ref{fig:levels}).  Due to the
  harmonicity within the Stark manifold, the states on this diagonal ladder can
  be interpreted as the $\ket{J,M}$ states of a large spin-$J$ system with
  $J=(n-1)/2$ and $M=m-J$ \cite{Facon2016}. The energy levels are pairwise
  separated by $\omega_\text{at}$, which is given by the first order Stark shift
  (cf.  Eq.~\eqref{eq:DeltaE_DC1}) and of the order of 100~MHz for $\mathcal
  E_\text{DC}=1$~V/cm and $n\approx 50$.  In this analogy, the circular state
  corresponds to the North pole $\ket{J,M=J}$ of the generalized Bloch sphere.
  The coordinates $(X,Y,Z)$ of the Bloch vector on the sphere are given by the
  expectation values of the $x$-, $y$- and $z$-component of the angular momentum
  vector $\vec{J}$, respectively. In these terms, applying
  a $\sigma^+$-polarised pulse with phase $\phi_\text{RF} = \varphi + \pi/2$ and
  detuning $\delta = \omega_\text{at} - \omega_\text{RF}$ leads to a rotation of
  the Bloch vector along
  \begin{equation}
    \vec{\Omega} = (- \Omega_R \, \sin \varphi,\,
                      \Omega_R \, \cos \varphi,\,
                      \delta)
    \label{eq:omega}
  \end{equation}
  with the Rabi frequency
  \begin{equation}
    \Omega_R = 3 \, \mathcal{E}_\text{RF} \, n\,, \label{eq:omega_r}
  \end{equation}
  where $\mathcal{E}_\text{RF}$ is the amplitude of the RF field. In the
  resonant case, $\delta = 0$, and when starting from the circular state, the RF
  field induces a rotation to the $\left( \vartheta, \varphi \right)$-direction
  where $\vartheta = \Omega_R T$. The resulting state, $\ket{\vartheta,
  \varphi}$, is called a spin coherent state (SCS) in analogy to the coherent
  states in quantum optics \cite{Glauber1963}.

  When alkali atoms are considered, the quantum defects perturb the harmonicity
  of the Stark manifold (cf.\ red energy levels in Fig.~\ref{fig:levels}). For
  small values of $\abs{m_\ell}$, states can be missing from the manifold if the
  energy shift due to their quantum defect is much larger than their Stark
  shift.  For the DC fields considered here, states are missing in vertical
  ladders with $\abs{m_\ell}\leq2$. This results in irregular offsets of the
  vertical ladders and anharmonic diagonal ladders.  This can be seen from the
  blue arrows on the left hand side in Fig.~\ref{fig:levels}, which are detuned
  from the transitions between the red energy levels.  For $\abs{m_\ell}=2$,
  only a single state is missing, and the harmonicity of the lowest diagonal
  ladder towards higher values of $m_\ell$ is almost preserved, while it is
  broken towards the other side.  This makes the lowest level of the
  $\abs{m_\ell}=2$ ladder the ideal starting point for
  circularization~\cite{Signoles2014}, i.e., for the transition to the circular
  state with $m_\ell=n-1$ and $\mu=0$.

  To optimize the electron's dynamics in the circularization process, we need to
  numerically compute the Hamiltonian. Thus, we have to calculate the matrix
  elements of the interaction Hamiltonian, $V = -\vec{d} \cdot
  \vec{\mathcal{E}}$, where $\vec{d}=q\vec{r}$ denotes the dipole operator and
  $\vec{\mathcal{E}}$ is either the DC or RF electric field,
  $\vec{\mathcal{E}}_\text{DC}$ or $\vec{\mathcal{E}}_\text{RF}(t)$. To this
  end, the matrix elements are split into their radial and angular parts. The
  latter can be expressed in terms of the Clebsch-Gordan
  coefficients~\cite{Bethe1977}. The radial part is calculated numerically,
  using Numerov's method~\cite{Numerov1924}.

  To speed-up the numerical calculations during the propagation and
  optimization, the total Hilbert space has to be reduced. For this purpose, we
  only take into account states that have a large dipole matrix element with at
  least one of the `pivotal' states on the lowest diagonal ladder.  As it turns
  out, it is sufficient to take the lowest and second lowest diagonal ladders
  into account during the calculations. However, all calculations were
  cross-checked by comparing to calculations with a Hamiltonian defined on
  a larger Hilbert space that contains all states of the $n$-manifold.

  Finally, the time-dependent Schr\"odinger equation,
  \begin{equation}
    \label{eq:TDSE}
    i\hbar \frac{\partial}{\partial t}\ket{\psi(t)} = H(t) \ket{\psi(t)},
  \end{equation}
  is solved numerically with the Chebychev propagator~\cite{Kosloff1994}, using
  the qdyn library~\cite{QDYN}. In the numerical implementation, the
  $\ket{\psi(t)}$ state is expanded in the eigenbasis of the Stark states in the
  presence of the DC field. The time-dependence in the Hamiltonian arises from
  the time-dependence of the RF field strength,
  $\vec{\mathcal{E}}_\text{RF}(t)$.

\section{Optimal control theory}
\label{sec:oct}

  The goal of this work is to drive the population from the initial state
  $\ket{\Psi(0)}$ towards the target circular state $\ket{\Psi_\text{tgt}}$ as
  fast and as accurately as possible. To this end, we employ quantum optimal
  control theory (OCT) \cite{Glaser2015,Reich2012}. The success of the desired
  state-to-state transfer  can be quantified by the following `cost'
  functional~\cite{Palao2003},
  \begin{align}
    J_T = 1 - \abs{\braket{\Psi(T)|\Psi_\text{tgt}}}^2.
  \end{align}
  It depends implicitly on the set of external control fields
  $\{\mathcal{E}_k\}$ via the final state $\ket{\Psi(T)}$, where $k$ labels the
  different control fields. To limit the pulse amplitude or to
  avoid lossy regions of Hilbert space, one can impose additional constraints $g
  \big[\{\mathcal{E}_k\}\, ,\{\ket{\psi_j(t)}\}\big]$ to the fields and to the
  states at intermediate times $\{\ket{\psi_j(t)}\}$, where $j$ runs over the
  basis states~\cite{Palao2003,Palao2008},
  \begin{eqnarray}
    J\big[\{\ket{\psi_j}\}, \{\mathcal{E}_k\}\big]
    &= J_T + \int_0^T dt \, g\big[\{\mathcal{E}_k\},\{\ket{\psi_j(t)}\}\big].
    \label{eq:functional}
  \end{eqnarray}
  Here, we seek to minimise the required field amplitude and restrict ourselves
  to one control field, $\vec{\mathcal{E}}_\text{RF} (t)
  = \mathcal{E}_\text{RF}^x (t) \, \vec{e}_x + \mathcal{E}_\text{RF}^y (t) \,
  \vec{e}_y$, where the $x$- and $y$-component are optimized independently from
  each other. Consequently, even when we start with a $\sigma^+$-polarized guess
  pulse, the polarisation is not necessarily conserved in the optimized field.
  Then, the constraints in Eq.~\eqref{eq:functional} can be chosen
  as~\cite{Palao2003}
  \begin{align}
    g \big[\vec{\mathcal{E}}_\text{RF}\big]
    = \frac{\lambda}{S(t)}
      \left( \vphantom{f^f} \vec{\mathcal{E}}_\text{RF} (t)
                          - \vec{\mathcal{E}}_\text{ref}(t) \right)^2\ ,
      \label{eq:ga}
  \end{align}
  with $S(t)\in [0,1]$ being a shape-function to smoothly switch the field on
  and off, and $\lambda$ being a weight. The reference field
  $\vec{\mathcal{E}}_\text{ref}(t)$ is usually the pulse from the previous
  iteration~\cite{Palao2003,Reich2012}.  In the  following, we use Krotov's
  method~\cite{Krotov1999} in its adaption to quantum dynamics~\cite{SomloiCP93,
  Sklarz2002, Palao2003, Reich2012} to carry out the optimization. It consists
  in a sequential, gradient-based algorithm, which proceeds by forward
  propagation of the initial state and backwards propagation of the target
  state, followed by an update of the field in order to match the two.  The
  specific update formula used here is found e.g. in
  Refs.~\cite{Palao2003,Reich2012}.

\section{Fast circularization with constrained amplitude and bandwidth}
\label{sec:exp_circ}

  We refer to circularization as transfer from the lowest lying state of the
  $m_\ell = 2$-ladder to the circular state. This `initial' state has to be
  prepared beforehand, starting with a rubidium atom in the ground
  $5\textrm{S}_{1/2}$  state~\cite{Signoles2017}. To this end, the rubidium
  atoms are optically excited using three laser fields in the presence of a weak
  DC field \cite{Facon2016}. The target state of the optical excitation,
  $\ket{51f,2}$,  can be precisely addressed due to its weak quantum defect.
  Afterwards, the DC field is slowly increased to its final value. This enlarges
  the spacing of the states in the Stark manifold, and the state $\ket{51f,2}$
  becomes adiabatically the lowest lying state of the $m_\ell=2$-ladder.  Now,
  the actual circularization can take place, where the circular state is reached
  via an $n$-level Rabi oscillation.

\subsection{$\pi$-pulse}
\label{sec:guess}

  \begin{figure}[tb]
    \centering
    \includegraphics{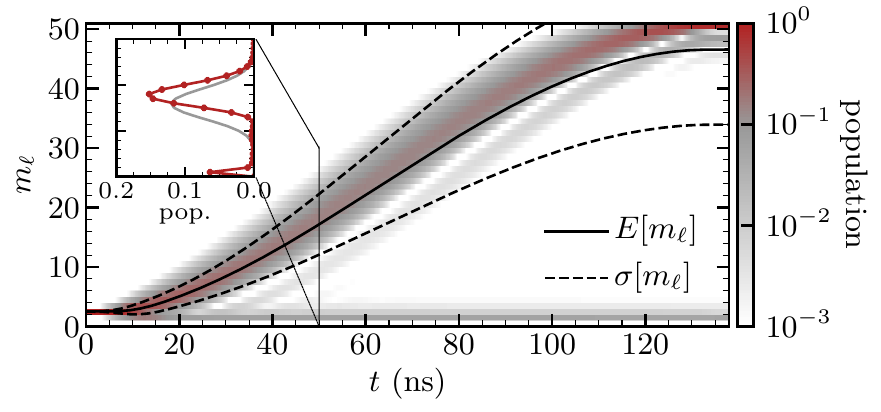}
    \caption{Distribution of the population over the lowest diagonal ladder for
      a $\pi$-pulse as described in Sec.~\ref{sec:guess}. Note the logarithmic
      scale in the color bar. The expectation value $E[m_\ell]$ and the
      standard deviation $\sigma [m_\ell]$ are indicated by the solid and dashed
      lines, respectively. The large variance is caused by the non-negligible
      population in $m_\ell = 1$. The inset displays the population distribution
      at $t=\ns{50}$ for $m_\ell \leq 30$. The grey line shows the population
      distribution of the closest SCS
      $\ket{\psi_\text{SCS}^{(\vartheta=0.62\pi,\varphi=0.49\pi)}}$.
    }
    \label{fig:guess_pop}
  \end{figure}

  As a  first guess for the optimization, we choose an RF $\pi$-pulse driving
  $\sigma^+$-transitions within the lowest diagonal ladder of the manifold. This
  pulse represents the present state of the art in the experiment
  \cite{Signoles2014,Signoles2017}.  Due to the quantum  defect, this pulse is
  near-resonant with transitions to the right side of the initial state, whereas
  it is slightly detuned from the transition to the left side (cf.
  Fig.~\ref{fig:levels}). Because only states in the lowest diagonal ladder play
  a significant role during the circularization, we will denote these states by
  $\ket{m_\ell}$ for the sake of simplicity. The initial state becomes $\ket{2}$
  and the target circular state $\ket{50}$.

  Following Ref.~\cite{Signoles2014}, we choose the $n=51$-manifold and
  $\mathcal{E}_\text{DC} = \Vpcm{2.346}$. The quasi-resonant frequency of the
  ladder is $\omega_0=\MHz{229.6}$,, which has been calculated using the
  first-order Stark splitting of hydrogen (cf.\ Eq.~\eqref{eq:DeltaE_DC1}). The
  chosen guess pulse is a $\sigma^+$-polarized RF pulse with central frequency
  $\omega_\text{RF}=\MHz{230}$ and amplitude $\mathcal{E}_\text{RF}
  = \mVpcm{18}$.  The pulse has a flat-top shape with sine-squared edges lasting
  $\ns{10}$ each. The total duration of the pulse, $t_\text{stop}=\ns{138}$, has
  been adjusted to match a $\pi$-pulse.

  The final population in the circular state is $81\%$ with an expectation value
  of $m_\ell$ equal to $46$. This indicates that the population is spread over
  several ladder states, as can be seen in Fig.~\ref{fig:guess_pop}.  A large
  fraction of the missing population ($6\%$ of the total population) can be
  found in the state $\ket{1}$. This effect can be attributed to an insufficient
  detuning of the RF pulse from the $\ket{2}\rightarrow\ket{1}$ transition. The
  transition frequencies $\omega_{m_\ell,m_\ell'} \equiv E_{\ket{m_\ell'}}
  - E_{\ket{m_\ell}}$ between the relevant states are
  \begin{subequations}
    \label{eq:critical_transitions}
    \begin{eqnarray}
      \omega_{0,1} &= \MHz{70.36}\,, \\
      \omega_{1,2} &= \MHz{182.95},\, \\
      \omega_{2,3} &= \MHz{227.46}.
    \end{eqnarray}
  \end{subequations}
  The quantum defect barely affects the $m_\ell=2$-ladder and $\omega_{2,3}$ is
  nearly resonant with the RF pulse, while the detuning from $\omega_{0,1}$ is
  large enough to prevent a significant population of state $\ket{0}$.  However,
  the rather small detuning from $\omega_{1,2}$ allows for an off-resonant drive
  towards state $\ket{1}$. While some of this population remains trapped in
  state $\ket{1}$, another part is reflected from the lower bound of the ladder
  and follows the main packet towards the circular state with a small delay.
  This can be seen from the light grey streak slightly below the main red path
  and from the strong deviation of the expectation value of $m_\ell$ from the
  center of the population distribution in Fig.~\ref{fig:guess_pop}.

  Moreover, the inset in Fig.~\ref{fig:guess_pop} displays the population
  distribution at the intermediate time, $t=\ns{50}$, together with the closest
  SCS $\ket{\psi_\text{SCS}^{(\vartheta=0.62\pi,\varphi=0.49\pi)}}$.  The latter
  is the SCS with coordinates that best match the expectation values of the
  three spatial components of the angular momentum vector for the system state
  $\ket{\psi(t=\ns{50})}$ (cf.  Sec.~\ref{sec:model}).  Apparently, the upper
  peak of $\ket{\psi(t=\ns{50})}$ is more narrow than a SCS and the residual
  population in the low-$m_\ell$ states affects the position of
  $\ket{\psi_\text{SCS}^{(\vartheta,\varphi)}}$ on the $m_\ell$-axis
  significantly. Thus, the overlap of
  $\ket{\psi_\text{SCS}^{(\vartheta=0.62\pi,\varphi=0.49\pi)}}$ and
  $\ket{\psi(t=\ns{50})}$ is only $65\%$.  As explained in Sec.~\ref{sec:model},
  a SCS can be rotated into the target circular state in a very natural way.
  However, due to the deviation from a perfect SCS, part of the population at
  the final time is spread over several states neighboring the target.

  To summarize our observations from Fig.~\ref{fig:guess_pop}, the fidelity
  obtained with a $\pi$-pulse is limited by (i) loss to $\ket{1}$, (ii) delay
  due to reflection from the lower end of the ladder, and (iii) an
  imperfectly-shaped SCS.  These effects together result in an infidelity of
  about $19\%$ at the final time. One could now directly use the $\pi$-pulse as
  a guess field to start the optimization, which provides an optimized pulse
  leading to a fidelity of $99\%$.  However, such a brute force approach comes
  at the expense of rather complex optimized fields with an undesirably large
  spectral bandwidth and a high field strength. In view of the experimental
  feasibility of the optimized pulse, it is much more advantageous to first
  exploit the available  physical insight and construct an improved guess pulse
  before starting the optimization.

\subsection{Two-step amplitude guess pulse}
\label{sec:newpulse}

  The observations above suggest to split the circularization into the
  preparation of a SCS and a subsequent rotation of the SCS into the target
  state.  Before constructing a pulse that implements such a two-step
  circularization, let us first estimate the maximal speed up of the
  circularization that can be expected.

  As with any dynamics on the Bloch sphere, the fastest rotation is obtained
  when the maximum allowed field amplitude is used. We choose
  $\mathcal{E}_\text{RF}=\mVpcm{45}$, which is close to the maximal,
  experimentally feasible value, $\frac{\Omega_R}{2\pi} = \MHz{9}$ or
  $\mathcal{E}_\text{RF}\simeq\mVpcm{46}$, cf. Eq.~\eqref{eq:omega_r}.
  Numerically, we find the optimal duration of the pulse to be $\ns{61.2}$ (data
  not shown), which is slightly faster than the analytical prediction for
  a $\pi$-pulse, which is $\ns{66.9}$.  The difference is explained by the the
  fact that the initial  state, $\ket{2}$, does not coincide with the South pole
  of the Bloch sphere.  Instead, it corresponds to a ring on the Bloch sphere
  around the $Z$-axis with $Z \approx -0.9$ (the radius of the sphere being
  normalized to $1$).

  While this amplified $\pi$-pulse is more than twice as fast as the one in the
  previous section, its fidelity is decreased to $63\%$. Obviously, for larger
  field strengths, the non-resonant run-off of population becomes more
  significant. In addition to the $\ns{61.2}$ for the mere rotation of a SCS,
  the accurate preparation of the SCS itself needs some time. Therefore, the
  approach to obtain high fidelity with the shortest possible duration is to
  increase the pulse duration gradually until the desired fidelity of $J_T
  = 10^{-2}$ can be reached under given constraints. Note that higher fidelities
  could easily be realized in the calculations. However, this will not be
  meaningful in view of application in an experiment where detector efficiency,
  DC electric field inhomogeneities, electric field noise and other experimental
  imperfections \cite{Facon2016} intrinsically limit the measurable fidelity.
  The lengthening of the pulse is motivated by the necessity to reduce the
  amplitude in the beginning of the circularization, in order to avoid
  population running off to the state $\ket{1}$ and allowing sufficient time for
  the preparation of the SCS.

  It turns out that the shortest possible pulse, cf.\ top panel of
  Fig.~\ref{fig:opt_pulse_pop} (grey shape), has a duration of $\ns{65}$. Its
  initial amplitude has been decreased to $\mVpcm{30}$.  After $\ns{10}$ the
  amplitude is increased to the maximum value, $\mVpcm{45}$. This increase takes
  another $\ns{10}$ and uses a sine-squared shape. The edges of the pulse are
  unchanged.  The new guess pulse, referred to as two-step amplitude pulse,
  leads to a fidelity of $74\%$, which is already a significant gain compared to
  the simple $\ns{61.2}$ $\pi$-pulse. Notably, the final population in state
  $\ket{1}$ under the new guess pulse  is decreased to only $6\times 10^{-6}$.

\subsection{Optimization results}
\label{sec:optresult}

  \begin{figure}[tb]
    \centering
    \includegraphics{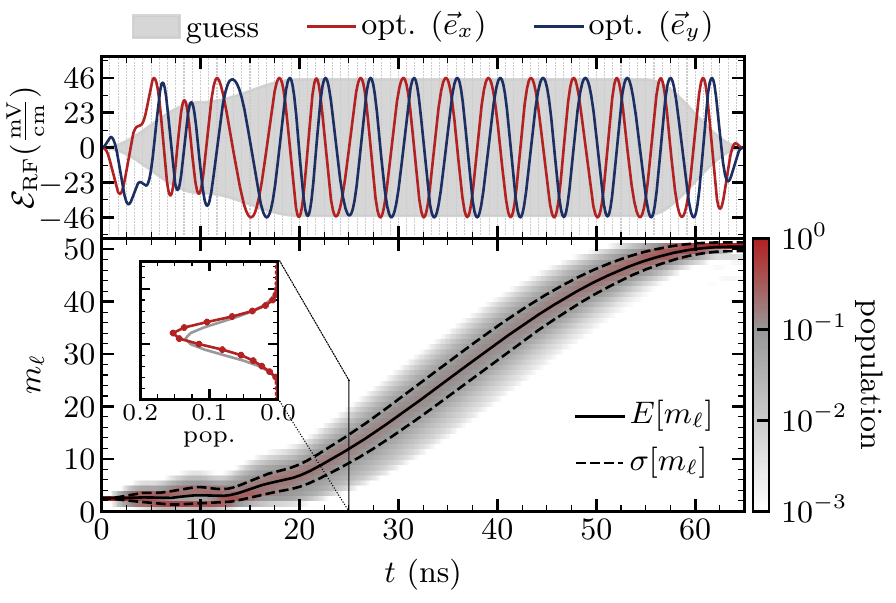}
    \caption{Top: Guess pulse (grey shape) and optimized pulse (red and blue
      lines corresponding to the $x$- and $y$-component of the pulse,
      respectively).  The dashed grid lines depict the step size of the coarse
      graining of the AWG ($\ns{0.83}$).
      Bottom: Same as Fig.~\ref{fig:guess_pop} but for the optimized pulse
      leading to a fidelity of $99\%$. The grey line shows the population
      distribution of the closest SCS
      $\ket{\psi_\text{SCS}^{(\vartheta=0.69\pi,\varphi=0.54\pi)}}$.
    }
    \label{fig:opt_pulse_pop}
  \end{figure}

  Next, we seek to increase the fidelity of the two-step amplitude pulse by
  employing Krotov's method (cf.\ Sec.~\ref{sec:oct}). The optimization is
  performed until the functional crosses the threshold $J_T = 10^{-2}$, which
  corresponds to a fidelity of $99\%$. The considered shape function $S(t)$
  (cf.\ Eq.~\eqref{eq:ga}) is the previously described sine-squared function. To
  ensure experimental feasibility, the absolute value of the amplitude is
  constrained to values smaller than $\mVpcm{46}$. Spectral components driving
  either $\sigma^-$-polarized or $\sigma^+$-polarized transitions above
  $\MHz{460}$ are suppressed.  This is most simply achieved by cutting the
  amplitude and spectrum of the pulse to the allowed range after each iteration.
  While such a procedure typically results in loss of monotonic convergence of
  the optimization~\cite{GollubPRL08,SchroederNJP09,LapertPRA09}, in our case
  $J_T$ does converge monotonically due to the high quality of the guess pulse.
  All in all, $1390$ iterations were needed to cross the threshold. One
  iteration requires a computation time of $\sim \unit[50]{s}$ on a standard
  work-station for the considered part of the manifold (the two lowest diagonal
  ladders with $m_\ell \geq 0$).

  Comparison of the guess and optimized pulse, cf.\ top panel of
  Fig.~\ref{fig:opt_pulse_pop}, reveals that the pulse is mainly changed in the
  low-amplitude step while it is left nearly unaltered in the high-amplitude
  step except for a shortening of the edge time. The shorter edge times and the
  stronger time dependence in the beginning of the pulse lead to a broadening of
  the spectrum but no conspicuous resonances appear in the spectrum other than
  the near-resonant transitions at $\MHz{230}$.

  Inspection of the population dynamics under the optimized pulse, cf.\ bottom
  panel of Fig.~\ref{fig:opt_pulse_pop}, reveals that the population is focused
  onto a few states with a small standard deviation and without any significant
  population remaining in the low-$m_\ell$ states. Moreover, it confirms the
  already predicted evolution: In the beginning, a SCS is generated, which is
  then driven towards the target state easily.  As can be seen in the inset, the
  state at $t=\ns{25}$ is very close to an SCS\@. Indeed, its overlap with the
  closest SCS $\ket{\psi_\text{SCS}^{(\vartheta=0.69\pi,\varphi=0.54\pi)}}$
  indicated by the grey line is $98.9\%$.

  \begin{figure}[tb]
    \centering
    \includegraphics{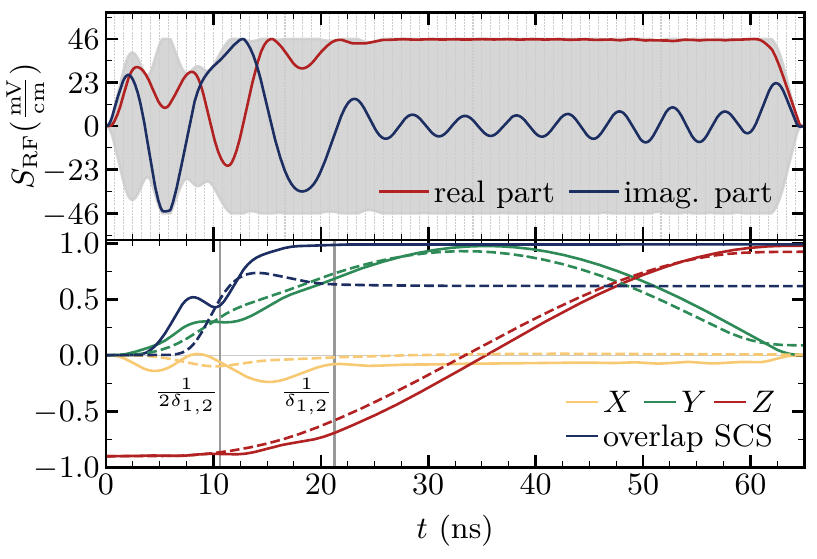}
    \caption{Top: Complex shape function of the optimized
      pulse. The pulse is separated in its real (red) and
      imaginary (blue) part while the envelope is represented by the grey shade
      in the background. The dashed grid lines in the background depict the step
      size of the coarse graining ($\ns{0.83}$).
      Bottom: Temporal evolution of the Bloch vector coordinates $X$, $Y$ and
      $Z$ in a frame rotating with $\omega_\text{RF}$ and of the overlap of the
      propagated state with the closest SCS
      $\ket{\psi_\text{SCS}^{(\vartheta,\varphi)}}$. Shown are the dynamics
      driven by the guess pulse (dashed lines) and optimized pulse (solid
      lines). The vertical lines show the time scale of the detuned
      $\ket{2}\rightarrow\ket{1}$ transition.
    }
    \label{fig:opt_shape_bloch}
  \end{figure}

  A deeper understanding of the dynamics is most easily gained when the process
  is considered in a frame rotating with $\omega_\text{RF}$. Accordingly, we
  demodulate the optimized pulse with the carrier frequency $\omega_\text{RF}$.
  The result is a complex envelope $S_\text{RF}(t)$ such that
  \begin{equation}
    \begin{multlined}
      \vec{\mathcal{E}}_\text{RF}(t)
        = \operatorname{Re}(S_\text{RF}(t))
           \big(\cos(\omega_\text{RF} t) \vec{e}_x
              + \sin(\omega_\text{RF} t) \vec{e}_y\big)\\
        + \operatorname{Im}(S_\text{RF}(t))
           \big(-\sin(\omega_\text{RF} t)\vec{e}_x
               + \cos(\omega_\text{RF} t) \vec{e}_y\big),
    \end{multlined}
  \end{equation}
  see also top panel of Fig.~\ref{fig:opt_shape_bloch}. The real part of the
  envelope can be interpreted as the shape of an RF pulse with phase
  $\phi_\text{RF} = 0$ (called the first quadrature), which induces a rotation
  around the $x$-axis of the generalized Bloch sphere (cf.\
  Eq.~\eqref{eq:omega}).  Conversely, the imaginary part corresponds to the
  shape of a pulse that is phase-shifted by $\pi/2$ (called the second
  quadrature), which induces a rotation around the $y$-axis.  The first
  quadrature  is the dominating one as can be seen from the red curve coinciding
  with the envelope of the pulse for most of the time.

  The role of the two quadratures becomes evident when considering the evolution
  of the state on the Bloch sphere, cf.\ bottom panel of
  Fig.~\ref{fig:opt_shape_bloch}.  It is constructive to consider the evolution
  driven by the guess pulse first (dashed lines).  The guess pulse consists of
  the first quadrature only, which induces a rotation around the $x$-axis. Since
  the initial state corresponds to a ring around the $z$-axis on the Bloch
  sphere, population with positive $Y$-components is rotated to higher
  $Z$-values whereas population with negative $Y$-components is rotated towards
  $\ket{1}$. This off-resonant transition needs $1/(2\delta_{1,2}) = \ns{10.6}$,
  which is additionally accompanied by a rotation around the $z$-axis (cf.\
  $z$-component of $\vec{\Omega}$ in Eq.~\eqref{eq:omega}). Because the
  rotation only affects part of the state, the cylindrical symmetry of the
  initial state is broken. This can be seen from the dip in the $X$-component at
  \ns{10.6}.  The accumulated population in $\ket{1}$ then oscillates back
  towards $\ket{2}$ at $1/\delta_{1,2} = \ns{21.2}$. Afterwards, the shape of
  the state does not change anymore (as can be seen from the constant overlap
  with a SCS) and the $X$-component is increasing only a little due to the
  second-order Stark shift. While, at the final time, the values of $X$, $Y$ and
  $Z$ are close to the target, the inaccurate shape of the state due to the
  dynamics in the first approximately \ns{20} is the main cause of the
  insufficient final fidelity.

  Thus, the main task of the optimized pulse is to improve the shape of the
  state. It is evident from Fig.~\ref{fig:opt_shape_bloch} that the first
  21$\,$ns of the pulse are designated to generate the required high-fidelity
  SCS. When the initial state splits in the very beginning, as explained above,
  both quadratures contribute to a clear separation of the two parts. This can
  be seen in Fig.~\ref{fig:opt_pulse_pop} from the partitioning of the
  population into two distinct branches around \ns{10}. In particular, the part
  with $m_\ell > 2$ is brought to a SCS-like shape much earlier, at \ns{8}, than
  for the guess pulse (cf.\ peak in the overlap of the state with a SCS in
  Fig.~\ref{fig:opt_shape_bloch}).  The $X$-value goes to zero here because the
  SCS-like state is located exactly opposite of the remaining population in
  $\ket{1}$.  When the population that has been driven to $\ket{1}$ joins the
  SCS-like state, the overlap decreases a bit before reaching its maximal value
  of $99\%$ after \ns{21.2}, in agreement with the time scale of the
  off-resonant oscillation to $\ket{1}$. Note that the remaining inaccuracy is
  retained until the end and is the main cause of the final error of $1\%$.

  The polar angle of the closest SCS
  $\ket{\psi_\text{SCS}^{(\vartheta,\varphi)}}$ amounts to $\vartheta = 0.75
  \pi$. The time that is required to rotate this state towards the North pole
  with the maximally allowed field strength is given by $T = \vartheta/\Omega_R
  = \ns{42}$. This rotation is visible in Fig.~\ref{fig:opt_shape_bloch} by the
  smoothly increasing $Z$-component and the flat shape of the first quadrature.
  At the same time, the second quadrature helps to adjust the $X$-component of
  the state to compensate for the drift due to the second-order Stark effect.
  The peak of the second quadrature at $\ns{62.5}$ induces a final adjustment of
  the $X$-component and kicks the SCS to the circular state. Finally, by adding
  a small overhead due to the edges of the pulse, the calculated time scales are
  in full agreement with the duration of the numerically determined time-optimal
  pulse.

  To corroborate our finding on ``time-optimality'', we have tested several
  values for initial and final pulse amplitude, duration and start of the
  amplitude ramp.  It turns out that the  crucial parameter for the success of
  the optimization under the given constraints is the pulse duration. This
  confirms the discussion of the previous paragraph. In fact, several different
  guess pulses with the same duration lead to almost the same optimized pulse.
  Pulses which are longer than $\ns{65}$ can also be optimized to the desired
  fidelity within a similar number of iterations (of the order of $2000$). On
  the other hand, no shorter pulse (such as the $\ns{61.2}$ pulse described
  above) can be optimized to the desired fidelity under the given constraints
  within at least $10^4$ iterations.

\section{Robustness against noise}
\label{sec:noise}

  To estimate the performance of our theoretical predictions under experimental
  conditions, we investigate the stability of the optimized two-step amplitude
  pulse with respect to different sources of noise.

\subsection{Coarse graining}
\label{subsec:coarse}

  In an experiment, the optimized pulse is realized by an arbitrary waveform
  generator (AWG), which samples the pulse at discrete points in time. The AWG
  is limited to a frequency bandwidth of $\MHz{480}$ for two channels. This
  constraint is fulfilled by the optimized pulse due to the bandwidth
  restriction during the optimization. In addition, the maximum sampling rate of
  $1.2\,$GS/s leads to a temporal step size of $\ns{0.83}$.  This limitation
  might have  a severe impact on the fidelity.

  To examine the influence of the limited resolution, we perform a coarse
  graining of the optimized pulse. In between two time steps, the pulse shape is
  interpolated linearly. This is a reasonable approximation of the real behavior
  of the pulse between two time steps.  A piecewise-constant approach has also
  been tested and leads to almost the same results.

  The step size of the coarse graining is indicated by the vertical dashed lines
  in the background of Fig.~\ref{fig:opt_pulse_pop} (top). In particular, at the
  maxima and minima of the pulse, there will be a significant difference between
  the original and the coarse-grained pulse. As a result, the increase in the
  $m_\ell$ expectation value is too slow, which results in a final fidelity of
  only $22\%$ while the most populated state is $\ket{49}$ (data not shown).
  Nevertheless, qualitatively the evolution is similar to the original one and
  it may be possible to attenuate the effect of coarse graining by increasing
  the amplitude in a well-adjusted way.

  A more elegant way to solve this problem consists in shaping directly the two
  quadratures. This corresponds to an optimization in the rotating frame and is
  shown in Fig.~\ref{fig:opt_shape_bloch} (top). It essentially circumvents the
  issue of limited time resolution altogether since the time dependence of the
  complex shape function $S_\text{RF}(t)$ is sufficiently slow to be insensitive
  to the coarse graining.  Thus, the fidelity decreases by only $0.06\%$
  compared to the original pulse without coarse graining. This is a striking
  improvement compared to the brute force approach and highlights once more the
  importance of choosing the proper frame of reference.

\subsection{Fluctuating field strengths}

  The second most important source of noise are fluctuations in the DC and RF
  fields. In theory, the DC field is assumed to be perfectly static, which is
  not the case in a real experiment. DC field fluctuations lead to an
  uncertainty in the position of the energy levels in the Stark manifold.  To
  check the stability of the pulse with respect to this noise, we have
  calculated the Hamiltonian for DC field strengths of $\mathcal{E}_\text{DC}
  \pm \Delta \mathcal{E}_\text{DC}/2$ with $\Delta \mathcal{E}_\text{DC}
  = \unit[50]{\upmu V/cm}$ and $\unit[150]{\upmu V/cm}$, corresponding to the
  best and worst case scenarios in a real experiment, respectively. Then, we
  have propagated the initial state under the action of each of these
  Hamiltonians and of the optimized two-step amplitude pulse. It turns out that,
  in both cases, the loss of fidelity with respect to the original DC field
  strength is of the order of $10^{-4}$ only.  The robustness of our optimized
  pulse can be understood as follows.  An offset in the DC field leads to the
  accumulation of an erroneous phase. This phase is proportional to the DC field
  offset and the pulse duration.  The  high stability of the optimized pulse is
  thus a direct result of its short duration.

  \begin{figure}[tb]
    \centering
    \includegraphics{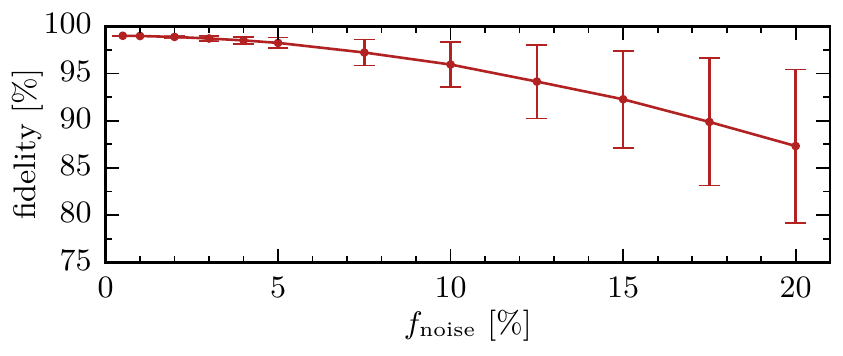}
    \caption{Fidelity of the optimized pulse as a function of noise strength,
      $f_\text{noise}$. Shown are the averaged values from $1000$ repetitions
      with randomly generated noise. The error bars indicate the standard
      deviation $\sigma$. The red line serves as a guide for the eye.
    }
    \label{fig:noise}
  \end{figure}

  Next, we consider amplitude noise in the RF pulse. To investigate its
  influence, we simulate fluctuations by adding a white-noise stochastic
  contribution. The noisy amplitude $\widetilde{\mathcal{E}}_\text{RF}(t)$ is
  realized as
  \begin{align}
    \widetilde{\mathcal{E}}_\text{RF}(t) = (1+R \cdot f_\text{noise})
               \mathcal{E}_\text{RF}(t)\,,
    \label{eq:noise}
  \end{align}
  where $R \in [-1,1]$ is a random number and $f_\text{noise} \in [0,1]$ is the
  noise strength. Note that the range of $f_\text{noise}$ implies amplitude
  fluctuations of at most up to a factor of two with respect to the original
  pulse. To avoid discontinuities, the random number $R$ is fixed for
  a half-period, i.e.\ between two zeros of the pulse. In other words, we assume
  the fluctuations to occur on a time scale slower than half a period of the RF
  pulse ($\sim \ns{2}$). Moreover, the $x$- and $y$-component are modified
  independently such that we also account for polarisation noise in the
  simulation.

  The impact of the RF amplitude noise on the fidelity of the target state is
  evaluated by repeating the propagation $1000$ times, each with a different
  realization of Eq.~\eqref{eq:noise}, for noise strengths between $0.5\%$ and
  $20\%$. As can be seen from Fig.~\ref{fig:noise}, the fidelity decreases for
  higher noise strengths and the spread becomes larger. For example, a fidelity
  of $95.9\%$ is expected on average for a noise level of $10\%$. This roughly
  corresponds to the present estimate of amplitude fluctuations in the
  experiment~\cite{Signoles2014,Signoles2017}, a value that could be reduced
  with simple experimental improvements.  Even at a level of $10\%$ amplitude
  noise, the  optimized pulse leads to significantly faster and more accurate
  circularization than achieved with e.g.
  a $\pi$-pulse~\cite{Signoles2014,Signoles2017}.

\section{Quantum speed limit}
\label{sec:qsl}

  Quantum mechanics itself sets a natural lower  bound on the minimal time that
  is needed to realize a certain dynamics~\cite{Giovannetti2003}. In the
  following, we will investigate the quantum speed limit for the circularization
  beyond (present) experimental constraints using optimal control theory. The
  speed limit has been investigated for multi-partite systems such as entangling
  gates between atomic qubits~\cite{GoerzJPB11} or transport in a spin
  chain~\cite{Caneva2009}. In these cases, it is the interaction strength
  between the subsystems that sets the speed limit. Here, such a limit is
  absent.  To the best of our knowledge, this is the first time that the speed
  limit for unary system dynamics in a realistic model is considered.

  \begin{figure}[tb]
    \centering
    \includegraphics{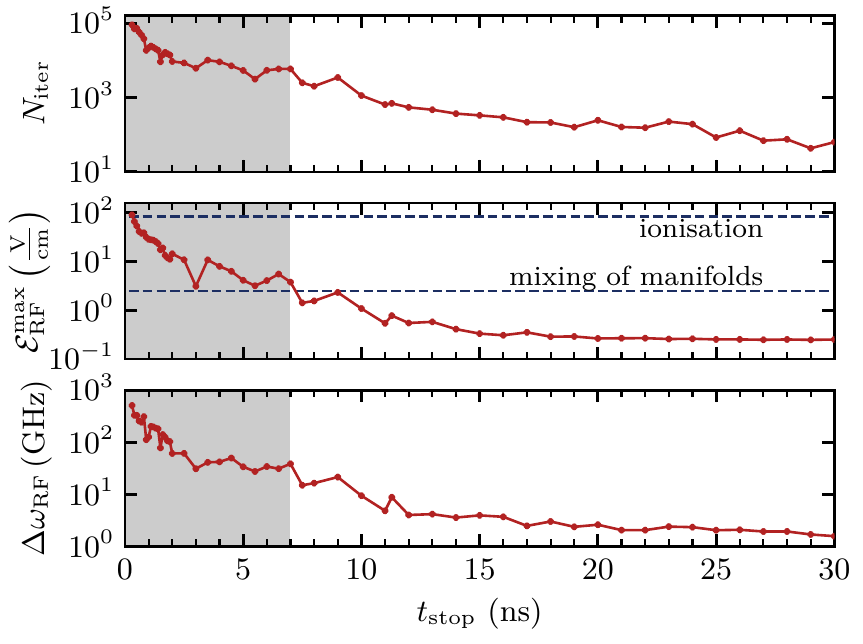}
    \caption{Number of required iterations $N_\text{iter}$ to reach a fidelity
      of $J_T = 10^{-2}$ together with the required maximal field strength
      $\mathcal{E}_\text{RF}^\text{max} = \max_t \abs{\mathcal{E}_\text{RF}(t)}$
      and bandwidth $\Delta \omega_\text{RF}$ of the corresponding optimized
      pulse for different pulse lengths $t_\text{stop}$.  The two field
      strengths where the adjacent manifolds start to mix with the central
      $n$-manifold and where the atom is ionized are marked by horizontal dashed
      lines. The grey background indicates the region where the model ceases to
      be predictive. The red lines serve as a guide for the eye.
    }
    \label{fig:qsl}
  \end{figure}

  To speed up the circularization, we gradually increase the field strength of
  the guess pulse.  In this context, it is important to keep in mind that our
  model consists of only the two lowest diagonal ladders. Eventually, this
  approximation will cease to be valid. We first discuss the QSL for our model
  and then examine the implications for the actual atom.  No constraints on the
  optimized pulse are included except for Eq.~\eqref{eq:ga}. The optimization is
  performed until the functional crosses the threshold $J_T = 10^{-2}$. For each
  value of $t_\text{stop}$, we have tested several guess pulses with different
  edge times and field strengths. In Fig.~\ref{fig:qsl}, we have plotted the
  number of iterations required to reach the desired fidelity for the best
  guess, i.e., the pulse that needs the least number of iterations. It is
  evident from the figure that the number of iterations increases as the pulse
  becomes shorter. The fluctuations in the data points can be credited to the
  high sensitivity of the optimization to the guess pulse.  In principle,
  Fig.~\ref{fig:qsl} suggests that it is possible to circularize the atom within
  less than  $\ns{1}$. However, when the field strength of the optimized pulse
  increases dramatically, our model will cease to be predictive for the actual
  atomic system since we account for a small part of the Hilbert space only.

  The first approximation is to neglect higher diagonal ladders in the
  $n=51$-manifold. Because the transition frequencies between the pivotal states
  and the omitted states are larger and the dipole matrix elements smaller than
  between the states in the considered part of the Hamiltonian,
  a re-optimization to reach the desired fidelity is possible in principle.  The
  optimized fields will certainly look different but the possibility of reaching
  the target state will not be affected.  Nevertheless, the numerical
  calculations become rather demanding.  One iteration of a pulse with
  $t_\text{stop} = \ns{7}$ needs a computation time of about $\unit[6]{s}$ in
  our model (involving $101$ states). In contrast, when all states of the
  central $n$-manifold are considered ($\sim 2500$ states), one iteration of the
  optimization requires approximately ten minutes. The total computation time is
  given by that estimate times the number of iterations, of the order of $10^4$
  for $t_\text{stop} = \ns{7}$, and increasing for decreasing $t_\text{stop}$.

  The next approximation is omission of adjacent Stark manifolds.  As the
  driving RF field becomes stronger, the states in the Stark manifold get
  increasingly dressed and the energy spacing increases. Once the dressing is
  strong enough to induce a crossing of adjacent manifolds, the instantaneous
  eigenenergies of our model will significantly differ from the ones obtained
  when considering the full atomic system. As a result, our model does not
  describe the actual atom any more. In the following, we will derive an
  approximation for the critical RF field strength where this effect becomes
  non-negligible.

  For reasons of simplicity we will neglect the quantum defect here. Because the
  position of the centre of a manifold is given by the Rydberg formula (cf.\
  Eq.~\eqref{eq:E_nl}) the central $n$-manifold will cross the upper
  $(n+1)$-manifold first. Moreover, the eccentricity quantum number takes its
  maximal possible value of $\abs{\mu} = n-1$ at the top and the bottom of the
  $m_\ell = 0$-ladder, which results in the largest possible first-order Stark
  splitting (cf.\ Eq.~\eqref{eq:DeltaE_DC} and Fig.~\ref{fig:levels}). In
  addition, the energy levels get dressed by the RF field.  The eigenstates of
  a perfect spin-$J$ system coupled to a field mode are separated in energy by
  the Rabi frequency $\Omega = \abs{\vec{\Omega}}$ (cf.\ Eq.~\eqref{eq:omega})
  \cite{Signoles2014}. When each pair of states acquires an additional energy
  splitting of $\Omega$, the edges of the diagonal spin ladder are shifted by
  $\Omega (n-1)/2$ with respect to the middle of the ladder. Adding up all
  contributions and considering a resonant RF field, the position of states at
  the top ($+$) or bottom ($-$) of the $m_\ell = 0$-ladder is given by
  \begin{equation}
    E_n^\pm = -\frac{1}{2n^2}
              \pm \frac{3}{2} (n-1)n \, \mathcal{E}_\text{DC}
              \pm 3 n \, \mathcal{E}_\text{RF} \, \frac{n-1}{2}.
  \end{equation}
  Solving $E_{51}^+ = E_{52}^-$ for $\mathcal{E}_\text{RF}$ with
  $\mathcal{E}_\text{DC} = \Vpcm{2.346}$ results in a critical field strength of
  \Vpcm{2.5} for the driving field. This value is in agreement with the
  numerical calculation of the instantaneous eigenenergies of the $n=50$ to $52$
  manifolds for different RF field strengths when taking the quantum defect into
  account.

  As a result, our model ceases to be predictive for RF field strengths above
  this value, which, according to Fig.~\ref{fig:qsl}, concerns all  pulses that
  are shorter than approximately $\ns{7}$.  For these RF fields strengths, our
  model provides no longer an appropriate description of the Rydberg atom and it
  is not possible to draw any definite conclusion about the possibility to
  circularize the atom with sufficient confidence. While it may still be
  possible to reach the target circular state under these very strong and broad
  pulses, the numerical effort to check the hypothesis poses a serious obstacle.
  When considering the extended Hilbert space containing all states with $n=51
  \pm 1$ and $m_\ell \geq 0$ ($\sim 4000$ states), the computation time of one
  iteration increases to one hour. Thus, the optimization becomes numerically
  infeasible because more than $10^4$ iterations are necessary to provide the
  desired fidelity.

  Eventually, when the field strength of the RF field is increased even further,
  a hard physical bound is reached when the fields are strong enough to ionise
  the atom.  For states on the lowest diagonal ladder, the order of magnitude of
  the necessary field strength is given by the static field ionization
  threshold, $\mathcal{E}=1/(9n^4)$ \cite{Gallagher}, which is $\Vpcm{83}$ for
  $n=51$.

  In summary, we find the speed limit to be determined by the internal structure
  of the atom.  In other words, the speed limit is defined in terms of the
  spectrum of the field-free Hamiltonian~\cite{Giovannetti2003} but for real
  physical systems, this Hamiltonian is typically an idealization. For very
  strong external control fields, this idealization ceases to be valid. It is
  thus important to keep the assumptions on which the Hamiltonian is based in
  mind when determining the quantum speed limit.

\section{Conclusions}
\label{sec:concl}

  We have tackled the question of how to prepare a Rydberg atom both quickly and
  accurately in a circular state. The necessary angular momentum transfer is
  realized by a suitably-shaped RF pulse.  Using quantum optimal control theory,
  we have shown that the circularization can be sped up by a factor of three to
  take merely 65$\,$ns. Even more importantly, the fidelity can be boosted from
  about 80\% to 99\%. The pulse has been constructed such as to be compatible
  with the current experimental setup~\cite{Signoles2017} in terms of maximal
  field amplitude and spectral bandwidth. We have found that the time needed for
  the circularization is mainly determined by the experimentally available RF
  field strength.

  We have tested our optimized RF pulses for robustness against various sources
  of noise. Overall, we find the protocol to be surprisingly stable. This is due
  to the short duration and small bandwidth of the pulse. The most detrimental
  source of noise are amplitude fluctuations of the RF field. Fluctuations above
  about 5\% are found to compromise the circularization fidelity.  When
  considering improvements of the experimental setup, stabilization of the RF
  field amplitude should thus be a priority. The optimized pulse is currently
  being implemented in the experiment and the first results look very promising.

  Key to the successful derivation of an experimentally feasible shaped RF
  pulse was the careful construction of a pre-optimized guess field, exploiting
  previous insight into the circularization dynamics. In fact, in quantum
  optimal control theory, many solutions to a given control problem can
  typically be found.  Which of these solutions is obtained in numerical
  optimization is then determined by the guess with which the iterative
  algorithm is initialized. Starting with a $\pi$-pulse that approximately
  rotates the system state onto the North pole of the $J=(n-1)/2$ spin Bloch
  sphere~\cite{Signoles2017} results in optimized pulses that are unnecessarily
  strong and spectrally broad.

  In general, pre-optimization of the guess pulse is a worthwhile endeavor  to
  guide the optimization towards experimentally feasible pulses. It can be
  carried out in a semi-automatic way by parametrizing the field and determining
  the best parameters before handing over to a gradient-based
  search~\cite{GoerzEPJQT15}.  In the case of circularization, a numerical
  pre-optimization of the guess pulse turned out not to be necessary. Analysis
  of the solution obtained by optimizing a $\pi$-pulse revealed that the
  dynamics can be split into the preparation of a SCS and a rotation, with  the
  proper angle, of this SCS towards the North pole. While the rotation requires
  the largest possible field strength to proceed at maximum speed, the
  preparation of the SCS is hampered by a large field strength. The run-off of
  population to other levels during the preparation of the SCS explains why
  a simple $\pi$-pulse fails. An intuitive solution is thus to decrease the
  amplitude during the SCS preparation, while ramping it up for the rotation of
  the SCS.  With such a guess field, the optimization has to introduce only
  small adjustments that do not come with an increase of bandwidth or amplitude.

  Splitting the transfer to the circular state into two steps furthermore allows
  us to identify what limits the minimum time required for the circularization.
  The time scale for preparing a SCS is given by the detuning of the RF pulse
  from undesired transitions to states with lower angular momentum projection,
  whereas the time to rotate the SCS to the North pole is determined solely by
  the Rabi frequency. A further speed up would thus be possible by using (a)
  a larger DC field strength, which increases the detuning, and (b) a larger RF
  field strength, which allows for faster rotation.

  For the given DC field strength, increasing the RF field strength could, in
  principle, bring the circularization time down to about 10$\,$ns.  However,
  this would come not only at the price of large RF amplitudes but also of
  a much broader spectral bandwidth. Therefore, given the present experimental
  technology, circularization times much below 50$\,$ns do  not seem realistic.

  Of course, it is still interesting to investigate, from a theoretical point of
  view, the fundamental limit to the minimum time for circularizing a Rydberg
  atom. Previously, the quantum speed limit has been discussed for multi-partite
  systems where the interaction strength is the key
  factor~\cite{Caneva2009,GoerzJPB11}, whereas here we have considered dynamics
  in a unary system. We have found that, using quantum optimal control theory,
  the duration of the circularization can be decreased by yet another order of
  magnitude while maintaining sufficient accuracy. However, the control fields
  become so strong that eventually the theoretical model ceases to be reliable.

  Since any theoretical description of a physical system is always based on
  idealization and simplifying assumptions, this finding is not restricted to
  Rydberg atoms. In fact, in a unary system, it is the spectrum of the
  field-free Hamiltonian alone that determines the speed
  limit~\cite{Giovannetti2003}. One could thus be led to think that it is
  possible to push the minimum duration for a certain dynamics to arbitrarily
  small values by using control fields that are strong enough to distort the
  spectrum. However, such strong fields will eventually always result in loss
  processes and thus introduce an effective speed limit. It would be interesting
  to rigorously account for this phenomenon in the fundamental theory of the
  quantum speed limit.

  \begin{acknowledgments}
    We thank E.-K. Dietsche and A. Larrouy for discussions. Financial support
    from the European Union under the Research and Innovation action project
    ``RYSQ''(Project No. 640378) is gratefully acknowledged.
  \end{acknowledgments}


  %

\end{document}